\documentclass[twocolumn,aps,pra,showpacs,amsmath,amssymb,superscriptaddress]{revtex4-1}

\usepackage[usenames]{color}
\usepackage{amsmath,stackrel}
\usepackage{soul}
\usepackage{mathrsfs}

\usepackage{amssymb}		
\usepackage{graphicx}		
\usepackage{hyperref}		
\usepackage{braket}
\usepackage{float}

\let \vec \mathbf

\begin{document}

\title{Quenched Magneto-association of Ultracold Feshbach Molecules}

\author{Kirk Waiblinger}
\affiliation{Department of Physics, University of Colorado, Boulder, Colorado, USA}
\author{Jason R. Williams}
\affiliation{Jet Propulsion Laboratory, California Institute of Technology, Pasadena, California, USA}
\author{Jos\'e P. D'Incao}
\affiliation{Department of Physics, University of Colorado, Boulder, Colorado, USA}
\affiliation{JILA, University of Colorado and NIST, Boulder, Colorado, USA}

\begin{abstract}
	We study enhanced magneto-association of atoms into weakly-bound molecules near a Feshbach resonance using a quench preparatory stage. In anticipation of experiments with NASA's Cold Atom Laboratory aboard the International Space Station, we assume as a baseline a dual-species ($^{87}$Rb and $^{41}$K) gas in a parameter regime enabled by a microgravity environment. This includes subnanokelvin temperatures and dual-species gases at densities as low as 10$^8$/cm$^3$. Our studies
	indicate that, in such a regime, traditional magneto-association schemes are inefficient due to the weak coupling between
	atomic and molecular states at low-densities, thus requiring extremely long magnetic field sweeps. To address this
	issue we propose a modified scheme where atoms are quenched to unitarity before proceeding with magneto-association.
	This substantially improves molecular formation, allowing for up to $80\%$ efficiency, and within time-scales much 
	shorter than those associated to atomic and molecular losses. We show that this scheme also applies 
	at higher densities, therefore proving to be of interest to ground-based experiments as well.
\end{abstract}

\maketitle

\section{Introduction}

Ultracold molecules are today one of the physical systems most used to study a variety of physical phenomena, 
ranging from quantum information \cite{soderberg2009NJP,carr2009NJP,park2017Sci,ni2018CS,anderegg2019Sci}, 
to ultracold chemistry \cite{krems2008PCCP,quemener2012CR,hu2019Sci,valtolina2020NT}, to exploration of novel dipolar 
phases of matter \cite{ni2008Sci,danzl2010NP,ospelkaus2010Sci,ni2010NT,wang2011PRLa,wang2011PRLb}, to tests 
of variations of fundamental constants \cite{chin2006PRL,chin2009NJP,borschevsky2011PRA}. 
As a result, developing efficient techniques to produce such molecules is a highly sought after goal
\cite{kohler2006RMP,hanna2007PRA,tscerbul2010PRA,owens2016PRA,ding2017PRA,dincao2017PRA,giannakeas2019PRL}. 
Since most experiments using such molecules start from a gas of ultracold 
atoms, the central question is how to efficiently produce a dense sample of molecules while still keeping them
at ultracold temperatures. Magneto-association provides such a path and has been routinely implemented to a broad range
of molecular experiments to date 
\cite{cubizolles2003PRL,herbig2003Sci,regal2003NT,strecker2003PRL,strecker2003PRL,durr2004PRL,mark2005EPL}. 
In this scheme, atoms are exposed to a magnetic field, $B$, tuned near a Feshbach 
resonance, causing the $s$-wave scattering length, $a$, to go through a pole, and causing interactions 
to become extremely strong \cite{chin2010RMP}. Sweeping the magnitude of the $B$-field across the 
resonance will convert atoms adiabatically to weakly bound Feshbach molecules, existing for $a>0$. 
Feshbach molecules can then be further used to explore a variety of phenomena or can be used as an intermediate
state to form more deeply bound molecular species by using other association schemes like 
STIRAP~\cite{bergmann2015JCP}.

The efficiency of magneto-association is fundamentally controlled by the $B$-field sweeping rates, 
but also depends on the initial atomic densities and temperatures
\cite{cubizolles2003PRL,herbig2003Sci,regal2003NT,strecker2003PRL,strecker2003PRL,durr2004PRL,mark2005EPL}, 
as well as depending non-trivially on the microscopic details characterizing the interatomic interactions 
\cite{kohler2006RMP,hanna2007PRA}.
For NASA’s Cold Atom Laboratory (CAL), a multi-user facility 
aboard the International Space Station \cite{elliott2018MC,aveline2020NT}, 
the microgravity environment will provide experimental conditions vastly different 
than those achievable in ground-based experiments exploring magneto-association. 
Here, we will show that although the unique experimental conditions available at CAL favor high phase-space 
density, long interrogation times, and suppression of gravitation-sag, 
the ultralow atomic densities ($n=10^8$-$10^{11}$/cm$^{3}$) desired for various proposed experiments
will drastically affect the efficiency of magneto-association. 
At such low densities, as opposed to those typically found in ground-based experiments ($n=10^{12}$/cm$^3$ and higher), the 
required $B$-field sweeps to obtain a satisfactory efficiency are simply too slow, compromising the stability of the molecular sample 
against three-body
losses.
A similar scenario is also found near narrow Feshbach resonances where atomic and molecular states are naturally 
weak.
This issue is of particular importance for studies of dual-species atom interferometry where the formation heteronuclear Feshbach molecules \cite{dincao2017PRA} is of fundamental interest to mitigate 
some of the major sources of systematic errors for high-precision tests of fundamental physics \cite{williams2016NJP,PhysRevLett.125.191101,RevModPhys.90.025008}.

In order to optimize the formation of Feshbach molecules, we modify the traditional magneto-association (tMA)
scheme. Our scheme adds a preparatory stage where the $B$-field is changed abruptly (quenched) from off-resonance 
to on-resonance, and then is allowed to dwell in this regime while developing correlations. 
This strongly correlated state will now serve as the initial state for magneto-association, providing a much higher 
overlap to the desired final molecular states.
This scheme is similar to the one used for $^{85}$Rb in Ref.~\cite{klauss2017PRL} which not only provided
association of dimers but also Efimov trimers \cite{dincao2018PRL}.
We show that this scheme, which we defined as quenched magneto-association (qMA), 
substantially improves the efficiency of molecule formation in the ultralow density regime, 
while still allowing it to be performed within time scales much shorter than those associated with atomic and molecular 
losses. Our proposed qMA is also found to be superior to tMA at higher densities, thus being of interest for ground-based experiments. More generally, we conjecture that qMA can also 
be potentially useful for molecular association studies near narrow Feshbach resonances since it provides a scheme to produce
a strongly correlated state for association regardless of the strength of the resonance.
Keeping in mind the relevant case of $^{87}$Rb-$^{41}$K mixtures available at CAL, 
our manuscript is organized as follows. In Section \ref{sec:theory} we describe our theoretical model and
emphasize its main assumptions and approximations. Section \ref{sec:mag} details both tMA and qMA schemes 
and analyzes the important time scales associated with the atomic and molecular losses. In Section \ref{sec:results} we
present our results for association of $^{87}$Rb$^{41}$K Feshbach molecules 
and discuss the main advantages of qMA while verifying its fundamental differences
to tMA schemes.

\section{Theoretical Model}{\label{sec:theory}}

\begin{figure}[htbp]
    \centering
    \includegraphics[width=\columnwidth]{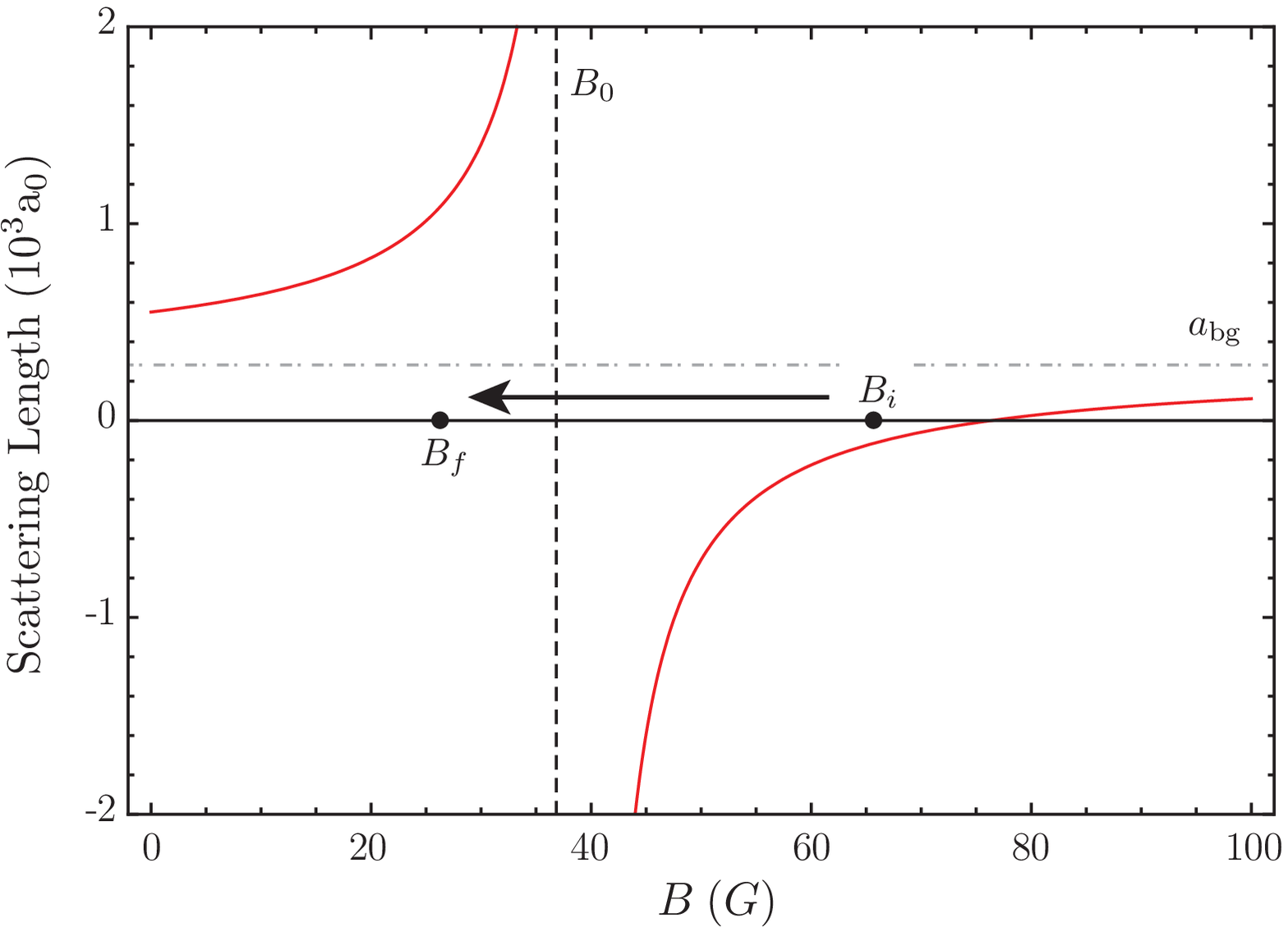}
    \caption{Scattering length (in units of the Bohr radius, $a_0$) as a function of applied $B$-field near the 
    $^{87}$Rb-$^{41}$K Feshbach resonance at $B_0\approx39.4$G. 
    This resonance is characterized by a width $\Delta B\approx37$G and background scattering length 
    $a_{\rm bg}\approx284$a$_0$~\cite{simoni2008PRA}.
    Both $^{87}$Rb and $^{41}$K atoms are in the $\ket{f=1,m_f=1}$ hyperfine state. $B_i$ and $B_f$ indicate, respectively,
    the initial and final values of the $B$-field in a hypothetical magneto-association scheme.}
    \label{fig:aVSb}
\end{figure}

In our present study we adopted two major assumptions that allow for a qualitative description 
of magneto-association while still providing a clear physical picture on
how medium (density) affects the association process. 
For the interatomic interactions we assume a single channel interaction model between 
$^{87}$Rb and $^{41}$K atoms, given by a Lennard-Jones potential
\begin{align}
    v(r)=-\frac{C_6}{r^6}\left(1-\frac{\lambda^6}{r^6}\right),\label{Vpot}
\end{align}
where $C_6\approx4274$a$_0^6E_{\rm h}$ \cite{derevianko2001PRA} 
is the van der Waals' dispersion coefficient and $\lambda$ is a tunable parameter adjusted to provide
the desired value of the scattering length. A more realistic description of the interactions between alkali 
atoms, however, is multichannel in nature and includes the hyperfine interactions responsible for 
the $B$-field dependence of the scattering length used in experiments with Feshbach resonances. A single 
channel description of this phenomena is supported by the universal properties of the system \cite{chin2010RMP} 
whenever $|a|\gg r_{\rm vdW}$, where $r_{\rm vdW}=({2\mu C_6}/{\hbar^2})^{1/4}/2$ is the van de Waals length 
and $\mu$ the two-body reduced mass. 
In our present study we model the $^{87}$Rb-$^{41}$K interactions near the well-known Feshbach resonance 
for atoms in the $\ket{f=1,m_f=1}$ hyperfine state (see Fig. \ref{fig:aVSb}) by adjusting the values of 
$\lambda$ in Eq.~(\ref{Vpot}) for each value of $B$ to produce the same value of $a$. As usual, for $B$-fields 
near a Feshbach resonance the scattering length is well represented by 
\begin{equation}{\label{eq:aofB}}
    a(B) = a_{\rm bg}\left(1-\frac{\Delta B}{B - B_0}\right),
\end{equation}
where the position of the resonance is $B_0\approx39.4$G, its width $\Delta B\approx37$G, and background scattering length
$a_{\rm bg}\approx284$a$_0$ \cite{simoni2008PRA}.

In order to incorporate density effects to properly describe magneto-association in ultracold quantum gases, 
we have employed the local density model 
\cite{borca2003NJP,goral2004JPB,stecher2007PRL,sykes2014PRA,corson2015PRA,colussi2018PRL}, 
allowing for a physically meaningful way to qualitatively describe the density dependence of various few-body 
observables. This model introduces a harmonic confinement to the few-body Hamiltonian,
whose strength is adjusted to produce a few-body ``density'' that matches that of the
experiment. In our current study, the two-atom Hamiltonian is then written as
\begin{equation}
    \hat{H} = -\frac{\hbar^2}{2\mu}\nabla^2+\frac{\hbar^2}{8 \mu a_{ho}^4} r^2 +v(r),\label{Ham}
\end{equation}
where $a_{\rm ho}$ is the harmonic oscillator length. We assume that the number 
densities of Rb and K are equal, i.e., $n_{\rm Rb} = n_{\rm K} = n$, and relate the oscillator length 
to the number density as \cite{sykes2014PRA}
\begin{equation}
 a_{\rm ho} = \sqrt{\frac{\pi}{8}}\left( \frac{4\pi n}{3} \right)^{-1/3}.
\end{equation}
This relation allow us to connect our few-body analysis with the relevant energy, length, and time scales
to the macroscopic system characterized, respectively,
\begin{align}
    E_n = \frac{(6 \pi^2 n)^{2/3}}{2\mu}\hbar^2,~~
    k_n = \frac{\sqrt{2\mu E_n}}{\hbar},~{\rm and}~
    t_n = \frac{\hbar}{E_n}.\label{nUnits}
\end{align}
We note that our model does not take into account quantum degeneracy and phase-space density
effects for association of Feshbach resonances as those experimentally observed in Refs. 
\cite{hodby2005PRL,thompson2005PRL,papp2006PRL}.
Nevertheless, the qualitative aspects of our analysis (in particular the comparison
between tMA and qMA protocols) still persists and should be observed in more 
elaborate models in which such collective effects are properly accounted for.

\begin{figure}[htbp]
    \centering
    \includegraphics[width=\columnwidth]{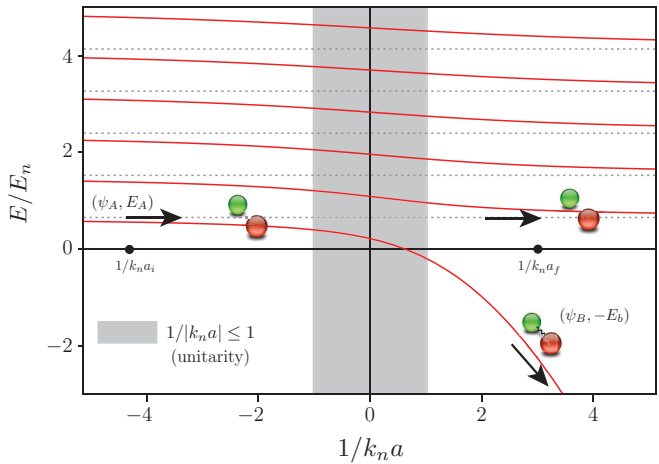}
    \caption{The energy levels of two atoms in a harmonic trap parameterized by $1/k_na$ and orbital angular momentum 
    $l=0$ (see text). In the limit of large ${1}/|{k_na}|$, the spectrum consists of pure harmonic oscillator levels,
    whose energies are plotted as dashed lines. Positive energies correspond to atomic states, and negative energies 
    represent molecular states. Note that the expected Feshbach molecular state for large positive~$a$, having binding energy 
    ${E_b={\hbar^2}/{2\mu a^2}}$, is shifted slightly due to the oscillator potential. The gray region indicate the values
    in which the system is found in the unitary regime, $1/|k_na|<1$.}
        \label{fig:eigenSpectrum}
\end{figure}

As a result, within our model magneto-association can be easily visualized through the two-atom energy spectrum 
as shown in Fig.~\ref{fig:eigenSpectrum}. The horizontal energy levels in the $1/|k_na|\gg1$ regime correspond to harmonic 
states and represent non-interacting atoms states. The desired target Feshbach molecular state is that with 
energy given by $-E_b=-\hbar^2/2\mu a^2$ for $a>0$ and is also indicated in Fig.~\ref{fig:eigenSpectrum}. 
[The energy spectrum is obtained by solving the two-body Schr\"odinger equation for the Hamiltonian in 
Eq.~(\ref{Ham}) for the lowest angular momentum state, $l=0$.]
The effect of sweeping the $B$-field from high to low (from $B_i$ to $B_f$ in Fig.~\ref{fig:aVSb})
corresponds to sweeping $1/k_na$ from left to right (from $1/ka_i$ to $1/k_na_f$) in Fig.~\ref{fig:eigenSpectrum}.
Transitions from atomic to molecular states are stronger in the {\em interaction region} $1/|k_na|\leq1$, i.e., 
when interactions are unitary, $n|a|^3\geq1$.
Therefore, within our physical picture, magneto-association reduces to the problem of non-adiabatic crossing of 
energy levels (Landau-Zener) \cite{clark1979PLA}. In our case, multiple levels can participate in the
process. Nevertheless, the Landau-Zener (two-level) model still provides a qualitative interpretation of the
phenomena and can serve as a guide for understanding of the important parameters controlling molecular 
association. For instance, within the Landau-Zener model \cite{clark1979PLA} the probability of 
transitioning from an energy level $\epsilon_1$ to $\epsilon_2$ via a 
linear sweep, $\epsilon_1-\epsilon_2=\alpha_\epsilon t$, where $\alpha_\epsilon$ is the sweeping
rate, is given by
\begin{align}
    P_{LZ}=1-e^{-2\pi\Gamma}.\label{PLZ}
\end{align}
Here, $\Gamma=\epsilon_{12}^2/\hbar\alpha_\epsilon$ with $\epsilon_{12}$ being the coupling between states $\epsilon_1$ and 
$\epsilon_2$. Applying this picture to our case in Fig.~\ref{fig:eigenSpectrum}, where energies are given in units of 
$E_n$ (\ref{nUnits}), allows us to access important information. 
For instance, since efficient association is obtained for $\Gamma\gg1$, 
the sweeping rates are required to be $\alpha_\varepsilon\propto n^{4/3}$, which can
be too slow in the ultralow density regime of CAL. 
We will explore these issues next and provide an alternative approach
to circumvent this limitation. 

\section{Quenched Magneto-association}{\label{sec:mag}}

As already anticipated from the discussion on the previous section, tMA schemes, i.e., 
applying a linear $B$-field sweep across a Feshbach resonance, might be inefficient for the low density regime relevant to a microgravity environment. Here we detail tMA and we discuss an alternative scheme that overcomes its 
limitations, but that can also be applied to ground based-experiments. We also discuss and characterize 
atomic and molecular losses.

\subsection{Sweeps and Quenches}{\label{ssec:schemes}}

\begin{figure}[htb!]
    \centering
    \includegraphics[width=3.0in]{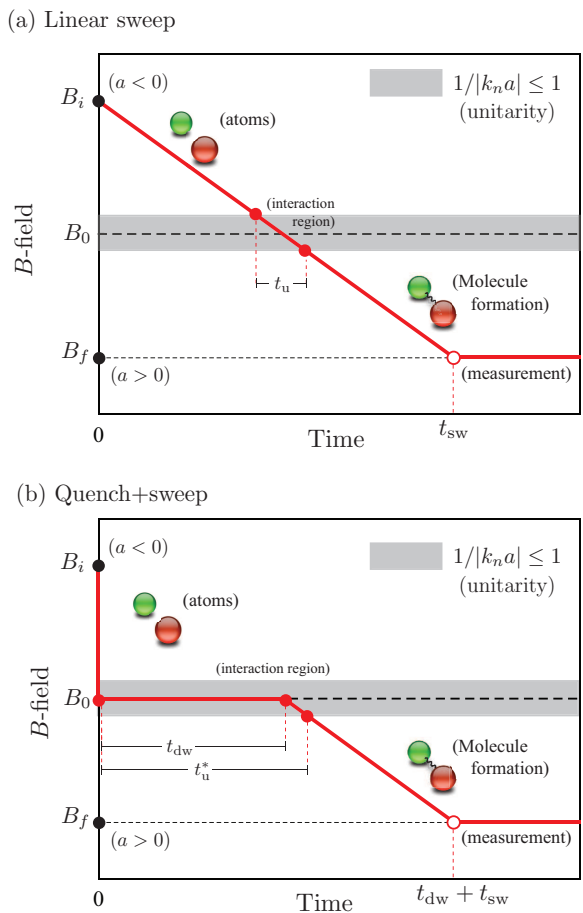}
    \caption{Schematic of magnetic field vs time in the traditional and quenched magneto-association schemes. 
    (a) In traditional magneto-association (tMA) $B$-field swept from $B_i$ to $B_f$ at a constant rate during a 
    time $t_{\rm sw}$. (b)     In quenched magneto-association (qMA) the $B$-field is instantaneously quenched 
    from $B_i$ to $B_0$, remaining at $1/|k_na|=0$ for a dwell time, $t_{\rm dw}$, 
    followed by a linear sweep from $B_0$ to $B_f$. This schematic figure is not to scale, and $B_i$ and $B_f$ 
    will not, in general, be equidistant from $B_0$, nor will $t_{\rm sw}$ and $t_{\rm dw}$ bear any particular 
    relationship to each other. In the figure, $t_{\rm u}$ and $t^*_{\rm u}$ represent the interaction time, i.e., 
    the time in which experience $1/|k_na|<1$, for tMA and qMA, respectively.}
    \label{fig:boft}
\end{figure}

Within our model, the key physical aspect that makes tMA inefficient at low
densities is the fact that during the $B$-field sweep the system remains in the interaction region, $1/|k_na|\leq1$,  
for a too short amount of time, thus requiring slow sweeps. 
The tMA scheme is illustrated in Fig.~\ref{fig:boft}(a), where the 
$B$-field is linearly swept from $B_i$ to $B_f$ during a time $t_{\rm sw}$.
In order to determine the interaction time, $t_{\rm u}$, we assume $B(t)=B_i-\alpha_B t$, 
where $\alpha_B=|B_f-B_i|/t_{\rm sw}\approx26.03{\rm G}/t_{\rm sw}$ is the sweep rate, 
and determine values of $B$ from Eq.~(\ref{eq:aofB}) in which the condition $1/|k_na|\leq1$ is satisfied. 
After some algebra one arrives at the interaction time given by
\begin{align}
t_{\rm u} \approx 2\left|\frac{(k_na_{\rm bg})\Delta}{\alpha_B}\right|\propto n^{1/3}t_{\rm sw}.\label{tu}
\end{align}
As a result, during a given sweep, atoms in the relevant states interact only during a much reduced 
amount of time as $n\rightarrow0$.

In order to improve on the interaction time, and consequently the association efficiency, we propose the scheme illustrated
in Fig.~\ref{fig:boft}(b). In such a scheme, the system is first quenched to $1/|k_na|=0$, corresponding to changing
the $B$-field from $B_i$ to $B_0$ within time scales much shorter than $t_n$. 
We note that at low densities the technical aspects of quenching the $B$-field becomes 
increasingly easier since $t_n$ increases as $n$ decreases. For our studies we assume that the quench is
performed instantaneously. After the quench, we allow the system to dwell for a time $t_{\rm dw}$ at $1/|k_na|=0$, 
thus letting interactions evolve before finally sweeping the field to its final value, $B_f$, 
accordingly to $B(t)=B_0-\alpha^*_{B}t$, where $\alpha^*_{B}=|B_f-B_0|/t_{\rm sw}\approx1.51{\rm G}/t_{\rm sw}$. 
As a result, in quenched magneto-association (qMA) the interaction time is now given by
\begin{align}
t_{\rm u}^*=t_{\rm dw}+\frac{t_{\rm u}}{2}=t_{\rm dw}+\underbrace{\left|\frac{(k_na_{\rm bg})\Delta}{\alpha_B^*}\right|}_{\propto n^{1/3}t_{\rm sw}},\label{tuS}    
\end{align}
which can be substantially enhanced by controlling $t_{\rm dw}$.
We note that this scheme is also similar to the one explored in Ref.~\cite{mark2005EPL}, though assuming 
$t_{\rm dw}\approx0$, and obtaining an efficiency of about 30\% at densities of $10^{12}$/cm$^{3}$.

Numerically, we study both tMA and qMA using the time propagation 
methodology developed in Refs.~\cite{dincao2018PRL,tolstikin2014}, with a few caveats introduced by the quench.
For tMA the initial state for propagation is a pure state, i.e.,
it is given by
\begin{align}
    \Psi_{i}\equiv\psi^{a=a_i}_{A}(\vec{r}),
\end{align}
where $\psi_{A}$ is an eigenstate of energy $E_A$ for $a=a_i$ (see Fig.~\ref{fig:eigenSpectrum}). 
In the quenched case, however, the initial state for propagation is instead a superposition of states given by
\begin{align}
    \Psi_{i}\equiv\sum_{\beta}c_{\beta}\:{\rm exp}\left[-\frac{iE_\beta t_{\rm dw}}{\hbar}\right]\:\psi^{a=\pm\infty}_{\beta}(\vec{r}),\label{PsiQ}
\end{align}
where $c_\beta=\langle\psi_A|\psi_\beta\rangle$, with $\psi_{\beta}$ and $E_\beta$ being the eigenstates
and energies of the system at $a=\pm\infty$. As we will see in Section \ref{sec:results},
the dependence on $t_{\rm dw}$ in this state plays a crucial role that improves the efficiency of  
magneto-association by letting interactions evolve at $1/k_na=0$.
Note also that for $t_{\rm dw}\gg t_n$ we expect that truly many-body effects to take place, 
potentially playing a rule in qMA. Current models do not not capture this physics properly, so we will
keep our study within modest values of $t_{\rm dw}$.
Before we compare in details both tMA and qMA schemes of Fig.~\ref{fig:boft}, 
we must first analyse the stability of the system with respect to losses, as
done in the next section.

\subsection{Atomic and molecular losses}{\label{ssec:losses}}

Regardless of the particular magneto-association scheme adopted, few-body losses 
can drastically reduce the efficiency of molecule formation. Although such loss processes are in general 
well understood in ultracold atomic and molecular gases \cite{dincao2018JPB}, magneto-association is a dynamical 
process and the full understanding on how losses occur as the interactions evolve is nontrivial.
In this section, however, we present an analysis that offers a qualitative understanding of the 
major loss processes, thus helping us to characterize and identify experimental regimes that are 
likely to mitigate their harmful effects.

In magneto-association the two major few-body loss processes are three-body recombination,
the process in which three free atoms collide to produce an atom and diatomic molecule,
and atom-molecule vibrational relaxation, causing a de-excitation of the molecular state. 
Both processes release enough kinetic energy to make their products to escape from typical traps \cite{dincao2018JPB}. 
In order to gain some insight on the time scales of the loss rates and their dependence on the experimentally
relevant parameters, we will consider the loss rates only in the regime at which they are maximal, i.e., $1/|k_na|<1$,
which is the regime most relevant to magneto-association. 
This analysis should provide an upper limit for the lifetime of the sample during the magneto-association 
process.

\begin{figure}[htbp]
    \centering
    \includegraphics[width=\columnwidth]{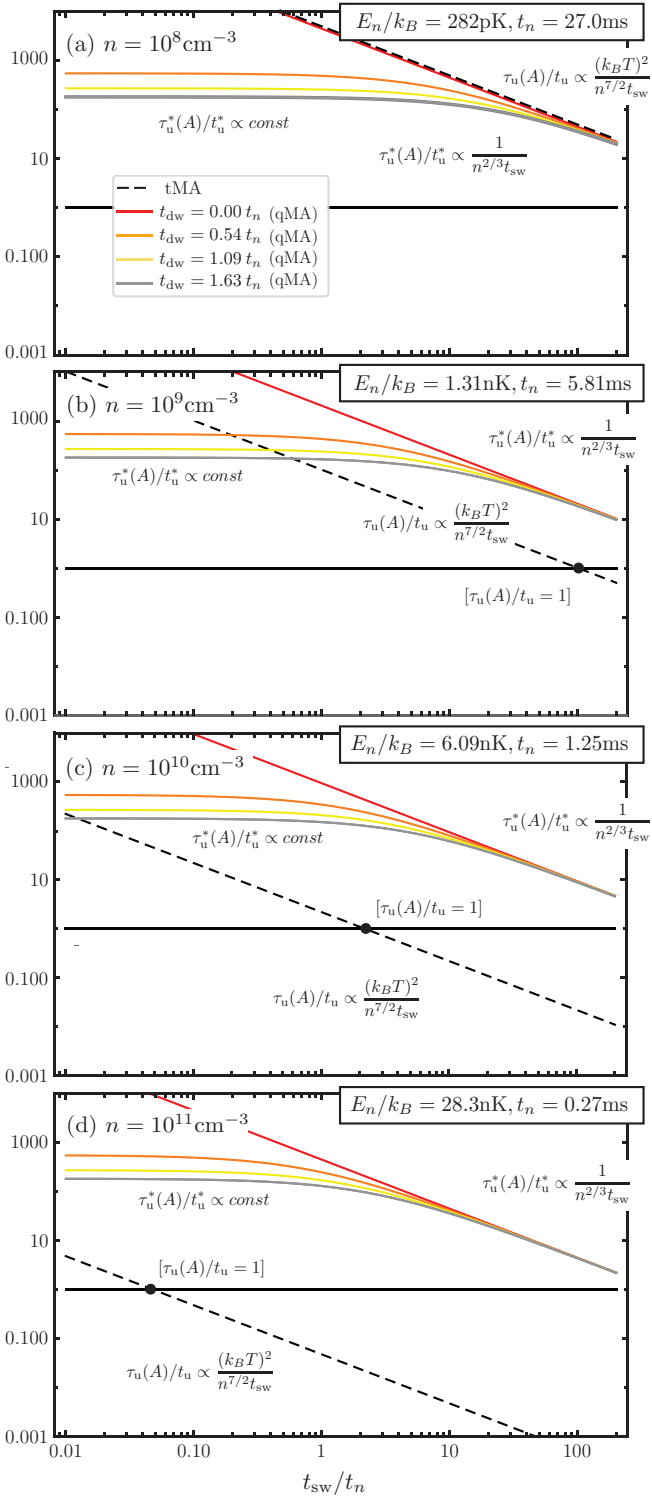}
    \caption{Analysis of the relevant time scales for magneto-association.
    For tMA, the values for $\tau_{\rm u}/t_{\rm u}$ calculated from the lifetimes due to 
    RbRbK losses at $T=100$pK (black-dashed curves) for various densities and a broad range of sweeping times. 
    As $n$ increases, the values of $t_{\rm sw}$ in which $\tau_{\rm u}/t_{\rm u}>1$ becomes quickly more
    restrictive. For qMA we display the values for $\tau^*_{\rm u}/t^*_{\rm u}$ (solid curves) for various values 
    of $t_{\rm dw}$. Regardless of the density, a broad range of $t_{\rm sw}$ satisfy the favorable condition for 
    magneto-association, $\tau^*_{\rm u}/t^*_{\rm u}>1$.}
        \label{fig:losses}
\end{figure}

It is well known that three-body recombination rate in the regime $1/|k_na|<1$ becomes independent of the
scattering length and estimated as \cite{dincao2018JPB}
\begin{align}
    L_{3}^{\rm u}(T)=\frac{4\pi^2\hbar^5}{\mu_{\rm 3b}^3(k_B T)^2}(1-e^{-4\eta}),\label{L3u}
\end{align}
where $\mu_{\rm 3b}^2=m_1m_2m_3/(m_1+m_2+m_3)$ is the three-body reduced mass, $T$ is the temperature, and
$\eta$ is the three-body inelasticity parameter, which provides a measure of the probability for inelastic 
transitions. The parameter $\eta$ is dependent upon the details of the interactions and in general obtained experimentally.
For the Rb-K mixture we are interested here, Ref.~\cite{barontini2009PRL} has determined $\eta\approx0.12$ for 
collisions involving 
two Rb atoms and one K (RbRbK), and $\eta\approx0.02$ for collisions involving two K atoms and one Rb (KKRb).
Now, assuming that the $B$-field sweep in tMA is performed at constant temperature,
the atomic lifetime is given by
\begin{align}
    \tau_{\rm u}(A)&=\frac{1}{n^2L_{3}^{\rm u}(T)}\nonumber\\
    &=\frac{9\pi^2(k_BT)^2\mu_{\rm 3b}^3}{8(1-e^{-4\eta})\mu^3\hbar^2}t_n^3\propto \frac{(k_BT)^2}{n^2},\label{TauT}
\end{align}
thus becoming shorter as the density increases and/or the temperature decreases. 
As a practical example, the lifetimes due to RbRbK losses at $T=100$pK and
densities of 10$^8$/cm$^{3}$, 10$^9$/cm$^{3}$, 10$^{10}$/cm$^{3}$, and 10$^{11}$/cm$^{3}$,
are, respectively, $37t_n$, $1.7t_n$, $0.08t_n$ and $0.0037t_n$, or, equivalently, 1s, 10ms, 
0.1ms and 1$\mu$s. In order to qualitatively understand what these lifetimes mean, the time scale of losses
needs to be compared to that of the interaction time [Eq.~(\ref{tu})], 
\begin{align}
    {\tau_{\rm u}(A)}/{t_{\rm u}}\propto\frac{(k_BT)^2}{n^{7/3}t_{\rm sw}}.
\end{align}
This indicates that an increase in the density, or a decrease in temperature, must be accompanied by a decrease 
of the sweeping time $t_{\rm sw}$ or, equivalently, an increase of the sweep rate $\alpha_B$, in order to compensate 
for the increase of atomic losses. As a result, since faster $B$\nobreakdash-field sweeps reduce efficiency, obtain a good balance
of losses and interction time, $\tau_{\rm u}/t_{\rm u}>1$, can only be done at the risk of compromising efficiency. 
This makes evident that finding the best regime for tMA is dependent upon a balance of various factors. 
Figure \ref{fig:losses} shows the values for $\tau_{\rm u}/t_{\rm u}$ calculated from the lifetimes due to 
RbRbK losses at $T=100$pK (black-dashed curves) for various densities and a broad range of sweeping times. 
Note that as $n$ increases, the values of $t_{\rm sw}$ in which $\tau_{\rm u}/t_{\rm u}>1$ becomes quickly more
restrictive. For each density, we have indicated the corresponding values for $E_n$ and $t_n$ in relevant units. 

In the case of qMA the key difference that improves the time scales for losses is that the quench 
itself increases the gas temperature and, according to Eq.~(\ref{L3u}), reduces the loss rates. Assuming 
that initial temperature is smaller than $E_n$, the quench sets the temperature to $k_BT=E_n$ \cite{makotyn2014NP}, 
thus leading to an atomic lifetime determined by
\begin{align}
    \tau_{\rm u}^*(A)&=\frac{1}{n^2L_{3}^{\rm u}(E_n/k_B)}\nonumber\\
    &=\frac{9\pi^2\mu_{\rm 3b}^3}{8(1-e^{-4\eta})\mu^3}t_n\propto \frac{1}{n^{2/3}}.\label{TauQ}
\end{align}

Interestingly, the lifetime now is linearly proportional to $t_n$ and consequently 
automatically rescaling for different densities and providing a weaker dependence on density than
Eq.~(\ref{TauT}). 
For instance, now the lifetimes due to RbRbK losses at $T=100$pK is 295$t_n$, which for
densities of 10$^8$/cm$^{3}$, 10$^9$/cm$^{3}$, 10$^{10}$/cm$^{3}$, and 10$^{11}$/cm$^{3}$,
become 8s, 1.7s, 370ms, and 80ms, respectively.
Comparing this lifetime to the interaction time in Eq.~(\ref{tuS}), we obtain in the
limit of long dwell times, $t_{\rm dw}\gg t_{\rm u}$,
an interaction time of $t^*_{\rm u}\approx t_{\rm dw}(\propto t_n)$, and 
\begin{align}
\tau_{\rm u}^*(A)/t^*_{\rm u}\propto const.
\end{align}
In the other hand, limit of short dwell times, $t_{\rm dw}\ll t_{\rm u}$
the interaction time is $t^*_{\rm u}\approx t_{\rm u}/2 (\propto t_n)$, and 
\begin{align}
\tau_{\rm u}^*(A)/t^*_{\rm u}\propto \frac{1}{n^{2/3}t_{\rm sw}}.
\end{align}
In either case, the above analysis indicates that qMA provides a substantially more favorable regime with 
respect to the scaling of losses with density and sweeping times.
Figure \ref{fig:losses} demonstrate this by displaying the values for $\tau^*_{\rm u}/t^*_{\rm u}$, also
calculated from the lifetimes due to RbRbK (solid curves), for various values of $t_{\rm dw}$. 
Note that regardless of the density, a broad range of $t_{\rm sw}$ satisfy the favorable condition for association,
$\tau^*_{\rm u}/t^*_{\rm u}>1$, a result much superior from those from tMA.

A similar analysis of the lifetime can be also provided once molecular association take places and
the relevant loss process is atom-molecule relaxation. Here, too, loss rates in the regime $1/|k_na|<1$ 
become independent on the scattering length and estimated as \cite{dincao2018JPB}
\begin{align}
    \beta_{\rm u}(T)=\frac{2^{1/2}\pi^{1/2}\hbar}{\mu_{\rm ad}^{3/2}(k_B T)^{1/2}}(1-e^{-4\eta}),
\end{align}
where $\mu_{\rm ad}=(m_1+m_2)m_3/(m_1+m_2+m_3)$ is the atom-molecule reduced mass. 
In this case, assuming that the $B$-field sweep in tMA is 
performed at constant temperature, the molecular lifetime is simply given by
\begin{align}
    \tau_{\rm u}(M)&=\frac{1}{n\beta_{\rm u}(T)}\nonumber\\
    &=\frac{3\pi^{3/2}\mu_{\rm ad}^{3/2}(k_BT)^{1/2}\hbar^{1/2}}{2\mu^{3/2}(1-e^{-4\eta})}t_{n}^{3/2}
    \propto \frac{(k_BT)^{1/2}}{n},
\end{align}
providing a weaker dependence on temperature and density than those for their atomic
counterparts in Eq.~(\ref{TauT}).
From the above equation, the molecular lifetimes due to RbRbK losses at $T=100$pK and
densities of 10$^8$/cm$^{3}$, 10$^9$/cm$^{3}$, 10$^{10}$/cm$^{3}$, and 10$^{11}$/cm$^{3}$,
are, respectively, $29t_n$, $14t_n$, $6.3t_n$ and $2.9t_n$, or, equivalently, 0.8s, 80ms, 
8ms and 0.8ms. These lifetimes are to be compared to that of the interaction time [Eq.~(\ref{tu})], 
\begin{align}
    {\tau_{\rm u}(M)}/{t_{\rm u}}\propto\frac{(k_BT)^{1/2}}{n^{4/3}t_{\rm sw}}.
\end{align}

In the case of qMA, assuming that initial temperature is smaller than 
$E_n$, the molecular lifetime determined by 
\begin{align}
    \tau_{\rm u}^*(M)&=\frac{1}{n \beta_{\rm u}(E_n/k_B)}\nonumber\\
    &=\frac{3\pi^{3/2}\mu_{\rm ad}^{3/2}\hbar}{2\mu^{3/2}(1-e^{-4\eta})}t_{n}
    \propto \frac{1}{n^{2/3}},
\end{align}
Here, the lifetimes due to RbRbK losses at $T=100$pK is 49$t_n$, which for
densities of 10$^8$/cm$^{3}$, 10$^9$/cm$^{3}$, 10$^{10}$/cm$^{3}$, and 10$^{11}$/cm$^{3}$,
become 1.3s, 0.3s, 62ms, and 13ms, respectively.
Comparing this lifetime to the interaction time in Eq.~(\ref{tuS}), we obtain in the
limit of long dwell times, $t_{\rm dw}\gg t_{\rm u}$,
\begin{align}
\tau_{\rm u}^*(M)/t^*_{\rm u}\propto const,
\end{align}
while for short dwell times, $t_{\rm dw}\ll t_{\rm u}$
\begin{align}
\tau_{\rm u}^*(M)/t^*_{\rm u}\propto \frac{1}{n^{2/3}t_{\rm sw}}.
\end{align}
The results above, lead to similar conclusions reached when analysing the atomic lifetimes.
Note that for the particular case of RbK mixtures, the typical molecular lifetimes $\tau^{\rm u}_M$ 
and $\tau^{\rm u*}_M$ are longer than those for the atomic lifetimes analysed above.

\begin{figure*}[htbp]
    \centering
    \includegraphics[width=6.8in]{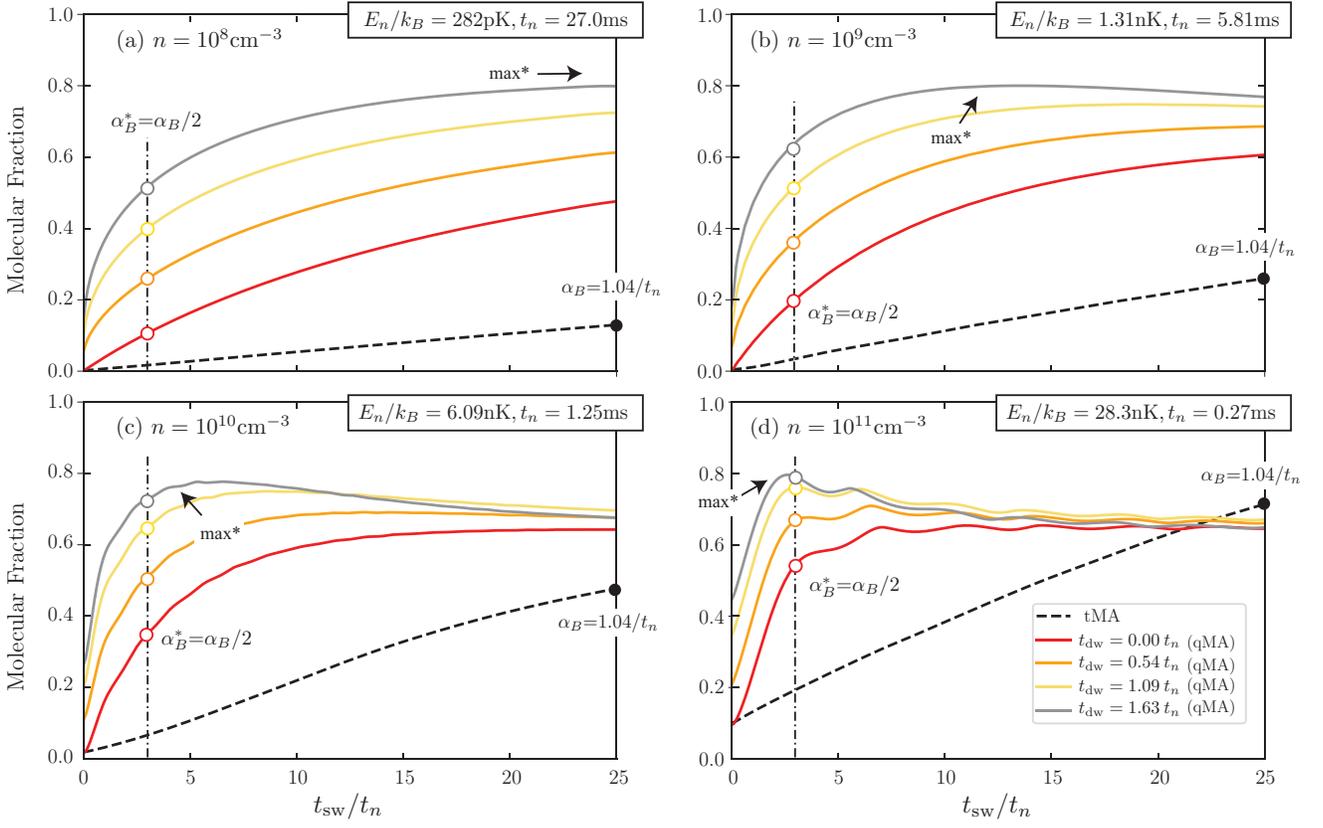}
    \caption{Fraction of molecules produced as a function of $t_{\rm sw}$ for various densities. 
    The dashed curve represent results for tMA while solid curves are those for qMA at different dweeling times, $t_{\rm dw}$. 
    In all cases the initial and final states are characterized by $a_i=-2r_{\rm vdW}$ and $a_f=100r_{\rm vdw}$, respectively.
    Note that at $t_{\rm sw} = 0$, tMA results are identical to those from qMA at $t_{\rm sw}=t_{\rm dw} = 0$. (See text for more discussions on these results and comparisons.) Figure display the corresponding values for $E_n$ and $t_{n}$.}
    \label{fig:resultsRealTime}
\end{figure*}

Overall, the analysis above indicates that the detrimental effects atomic on molecular losses to 
magneto-association are in general mitigated in the low-density regime, regardless of the magneto-association 
scheme. 
The CAL fully takes advantage of this property due to its unique capability to provide ultralow-density 
samples. 
The qMA, however, seems to provide a more stable scheme, in particular at higher densities,
due to its independence of temperature and the much more favorable ratio between lifetime and interaction
times, $\tau^{\rm u*}_A/t^*_{\rm u}$. The caveat of using quenches is that the final temperature of the
molecular sample will directly depend on the density (via its dependence on $E_n$), 
an effect that can once again mitigated at low-densities. Although such prospects are positive with 
respect to losses at the low-density regime, the question we now focus is how the actual efficiency of 
both magneto-association schemes compares to each other in this regime and whether association
occurs in time scales shorter than the atomic and molecular losses. We provide such an analysis 
in the next section.

\section{Molecular Association}\label{sec:results}

In this section we present our numerical simulation for both tMA and qMA near the $^{87}$Rb-$^{41}$K  Feshbach 
resonance at $B_0=39.4$~G~\cite{simoni2008PRA} (Fig. \ref{fig:aVSb}).
In each case, we select $a_i =$-2$r_{\rm vdW}$ ($\approx-$144a$_0$) and $a_f$=100$r_{\rm vdW}$ ($\approx$7230a$_0$), 
corresponding to $B_i= 63.92 $~G and $B_f= 37.89$~G. Therefore, we are assuming bosonic heteronuclear Feshbach 
molecules which are substantially larger than previously studied \cite{klempt2008PRA,weber2008PRA}
which aligns best with CAL's future experiments in microgravity \cite{dincao2017PRA}. As a result, most
of our studies presented here will be performed for atomic densities in which the average interatomic distances are 
larger than the molecular size, i.e., $n\le10^{11}$/cm$^3$.
We study the molecular fraction in terms of the sweeping time, $t_{\rm sw}$, dwell time, $t_{\rm dw}$, 
and atomic densities.
We note that the qualitative trends in our results do not depend on the particular choices of $B_i$ and $B_f$, 
so long as they are chosen off resonance, i.e. $1/|k_na|\gg1$.

Here, we chose to compute the molecular fraction for four different atomic densities $n$= 
$10^8$/cm$^{3}$, $10^9$/cm$^{3}$, $10^{10}$/cm$^{3}$, and $10^{11}$/cm$^{3}$, thereby covering the
density regime available the CAL environment. Results are displayed in Fig.~\ref{fig:resultsRealTime},
where dashed lines represent the molecular fractions obtained via tMA,
and solid lines the ones obtained via qMA for different values of $t_{\rm dw}$ 
[see color-coded legend on the inset of Fig.~\ref{fig:resultsRealTime}(d)]. For each density, we also have indicated
the corresponding values for $E_n$ and $t_n$ as they characterize the typical energies and time scales
of the system at a given density $n$. These calculations yield a few crucial observations, valid for 
all densities considered. 

First, as shown in Fig.~\ref{fig:resultsRealTime}, in the low-density regime qMA
produces a higher molecular fraction than tMA within the same value for $t_{\rm sw}$. 
This immediately implies the most significant result of this study: 
within a given $t_{\rm sw}$, a quench with finite $t_{\rm dw}$ followed by a $B$-field sweep away from unitarity will in general
produce a larger molecular fraction than a pure $B$-field sweep in the same amount of time. Therefore, in order 
to produce a larger fraction of molecules in the shortest amount of time, qMA is clearly the optimal choice.
The faster times for molecular formation within qMA is crucially important since it ensures the mitigation of atomic 
and molecular losses, in particular for higher densities, as discussed in Section.~\ref{ssec:losses} 
(see Fig.~\ref{fig:losses}). Figure \ref{fig:resultsRealTime} also indicates that qMA reaches an efficiency of 
about $80\%$, reaching this maximum value for smaller values of $t_{\rm sw}/t_n$ values as $n$ increases (see points in
Figure \ref{fig:resultsRealTime} marked by ``max*"). 
The finite efficiency in qMA is likely to be associated to the nature of the quenched state in 
Eq.~(\ref{PsiQ}), where excited states are more likely to remain in the atomic states than the target molecular 
state which the $B$-field is swept away from the interaction region $1/|k_na|<1$. A more complete analysis on this
topic is beyond the scope of our present study. 
For tMA, although one would in theory expect a nearly $100\%$ efficiency as $t_{\rm sw}\rightarrow\infty$, 
the results in Fig.~\ref{fig:resultsRealTime}, along with those for the atomic losses in Fig.~\ref{fig:losses},
clearly shows that substantial losses will take place well before being able to reach this level of efficiency.

In order to better understand the physical aspects determining the superiority of qMA over tMA 
we look at the values of the molecular fraction at $t_{\rm sw}=0$ in Fig.~\ref{fig:resultsRealTime}.
This value indicates the quality of the overlap between the initial state (prior the $B$-field sweeping) and the 
final molecular state. As shown in Fig.~\ref{fig:resultsRealTime}, while for tMA and qMA ($t_{\rm dw}=0$)
the molecular fraction are nearly identical, it increases for qMA with $t_{\rm dw}$. This is consistent with the 
experimental observations in Ref.~\cite{mark2005EPL} and demonstrates that the dwelling time plays an important role 
in the state preparation by letting correlations evolve in the interaction region $1/|k_na|<1$.
We note, however, that for $t_{\rm sw}>0$ the sweeping rates implied in Fig.~\ref{fig:resultsRealTime} are, 
in fact, different. While for the tMA the sweep rate is $\alpha_{B}=|B_f-B_i|/t_{\rm sw}\approx26.03{\rm G}/t_{\rm sw}$ 
in qMA the rate is $\alpha^*_{B}=|B_f-B_0|/t_{\rm sw}\approx1.51{\rm G}/t_{\rm sw}$.
The slower values for $\alpha^*_{B}$ thus partially explains the higher molecular fraction obtained via qMA for a 
given value of $t_{\rm sw}>0$. 

\begin{figure}[htbp]
    \centering
    \includegraphics[width=\columnwidth]{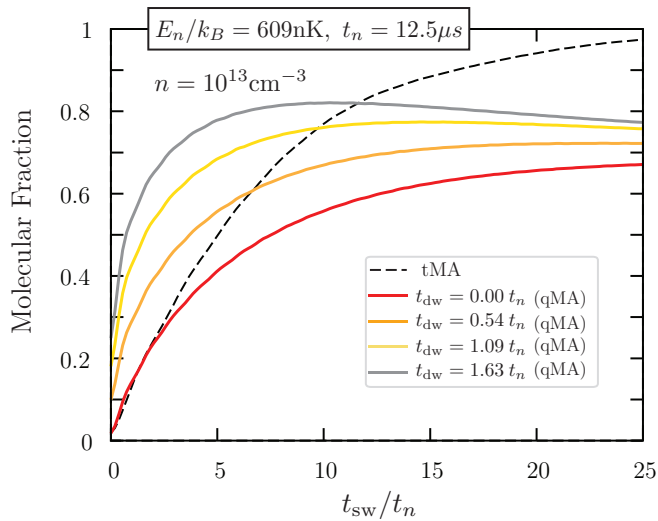}
    \caption{Fraction of molecules produced as a function of $t_{\rm sw}$ for $n=10^{13}$/cm$^3$. 
    The dashed curve represent results for tMA while solid curves are those for qMA at different dweeling times, $t_{\rm dw}$. 
    In all cases the initial and final states are characterized by $a_i=-2r_{\rm vdW}$ and $a_f=10r_{\rm vdw}$, respectively.}
    \label{fig:HighDens}
\end{figure}

In order to make a more direct comparison between the efficiency of both schemes we should compare, for instance, 
the molecular fractions obtained for sweeping rates that leads to the same sweeping times within the interaction 
region, $1/|k_na|=1$. Now, considering that for tMA one expands twice the time sweeping in this region than for 
qMA, we will seek for values of $t_{\rm sw}$ in which $\alpha^*_B=\alpha_{B}/2$. We note, however, since in tMA and qMA
the system experiences interactions differently during the time spent sweeping across the $1/|k_na|<1$, our comparison
here can only provides a rough analysis. Our comparison is shown in Fig.~\ref{fig:resultsRealTime}. We indicate 
the molecular fraction for tMA at $t_{\rm sw}=25t_n$ by a solid circle, producing the rate $\alpha_B=1.04/t_n$. 
This value should then be compare to those obtained via qMA at $t_{\rm sw}=2.90t_n$, as indicated by the open circles 
in Fig.~\ref{fig:resultsRealTime}. This analysis again demonstrate that for low densities, as $t_{\rm dw}$ increases,
the efficiency of qMA will surpass that of tMA, while still providing a much faster scheme to associate atoms
into molecules.

In order to illustrate the comparative performance of both association schemes at higher densities, 
and other typical parameters, Figure \ref{fig:HighDens} shows our simulations for tMA and qMA for an atomic density of $n=10^{13}$/cm$^3$.
This calculation associate atoms to a Feshbach molecular state with $a=10r_{\rm vdW}$ and 
demonstrate again the superiority of qMA in creating molecules at a faster rate. This indicates that qMA not only provides a vital
tool for molecular creation in microgravity, but also in typical regimes in ground-based experiments.

\section{Conclusion}

In this manuscript we have investigated methods to produce molecules via magneto-association near a Feshbach 
resonance, focusing on the low-density regime relevant to the microgravity environment of CAL. 
Based on the trends discovered from our computations, we conclude that a qMA can generally be made superior to tMA. 
Within qMA, we found that the dwelling time at the interaction region, $1/|k_na|<1$, allows for correlations
to develop, thus providing a much more efficient scheme for association of atoms in molecules. Our results
show that qMA allows for a much higher association efficiency ($\sim80\%$) within considerably faster 
time scales than tMA. This allows for further mitigation of atomic and molecular losses, regardless of the
density regime. 

In further studies of molecular production, several more complicated aspects of the system could be 
investigated. For instance, one could introduce dynamically the various loss processes dependent on scattering 
length, and attempt to optimize molecular production constrained by the loss timescales in a more rigorous way than 
presented here. Also, while the analysis in this study considered purely two-body interactions, one could extend 
the analysis by incorporating three-body effects, which would then have to consider both Rb-Rb-K and Rb-K-K 
interactions, as well as the formation of triatomic Efimov states existing in the system \cite{klauss2017PRL}. 

\acknowledgments

This research was carried out under a contract with the National Aeronautics and Space 
Administration. KW acknowledges support provided by the Undergraduate Research Opportunities Program 
(UROP) at the University of Colorado Boulder. JRW and JPD were supported by JPL-CALTECH under NASA contract NNH13ZTT002N. JPD also acknowledges partial support from the U. S. 
National Science Foundation, grant number PHY-2012125.


\end{document}